# Three-dimensional jamming and flows of soft glassy materials

G. Ovarlez, Q. Barral, P. Coussot

*Université Paris Est, Laboratoire Navier, LMSGC (CNRS-ENPC-LCPC), Champs sur Marne, France*

**Various disordered dense systems such as foams, gels, emulsions and colloidal suspensions, exhibit a jamming transition from a liquid state (they flow) to a solid state below a yield stress[1]. Their structure, thoroughly studied with powerful means of 3D characterization[2,3,4,5,6], exhibits some analogy with that of glasses[1,7,8] which led to call them soft glassy materials[9]. However, despite its importance for geophysical and industrial applications[10,11], their rheological behavior[2,12], and its microscopic origin[1,13], is still poorly known, in particular because of its nonlinear nature. Here we show from two original experiments that a simple 3D continuum description of the behaviour of soft glassy materials can be built. We first show that when a flow is imposed in some direction there is no yield resistance to a secondary flow: these systems are always unjammed simultaneously in all directions of space. The 3D jamming criterion appears to be the plasticity criterion encountered in most solids[14]. We also find that they behave as simple liquids in the direction orthogonal to that of the main flow; their viscosity is inversely proportional to the main flow shear rate, as a signature of shear-induced structural relaxation, in close similarity with the structural relaxations driven by temperature and density in other glassy systems.**

Although they have "solid" and "liquid" regimes, soft glassy materials present strong differences with usual solids and liquids that make the modelling of their behaviour challenging. E.g., their microscopic behaviour at the solid/liquid transition is very different from that of usual solids in which plasticity occurs through dislocations[14]. Their constitutive elements are trapped in cages from which they cannot spontaneously escape[2,3]; flow occurs when cage distortion by shear allows for particles rearrangements[5,15]. They also present strong differences with simple liquids: their liquid state is characterized by a complex nonlinear, usually shear-thinning, behavior[1,11], the origin of which is still a matter of discussion[9,16,17,18]; it may involve speeding up of structural relaxation by shear[4,16] and elastic coupling between rearrangement zones[18].

Most models, simulations and experiments consider the case of simple shear, which involves only scalars variables – the shear stress and the shear rate –. Generally 3D formulations of constitutive equations, essential for the description of complex flows in practical situations, cannot directly be inferred from these results. Besides, a better understanding of the structure of soft glassy materials and its link with flow properties might be gained by focusing on their 3D rheological properties. E.g., the 3D yield criterion is closely related to the shape of the energy landscape in which their elements are trapped, while the 3D resistance to shear at the neighbourhood of the jamming transition might provide an insight in their structural relaxation.

In this work, in the aim of investigating their 3D behaviour, we unjam soft glassy materials in a given direction and observe their behaviour in an orthogonal direction. We focus on three systems of different structures: an emulsion, a physical gel, and a colloidal gel (see Methods). With these systems, we have different physical origins of unjamming: droplet deformation (emulsion), blob squeezing (physical gel), contact breaking (colloidal gel), these different effects leading to collective rearrangements and flow.

We first use spheres of radius $R$ (much larger than the size of fluid elements) embedded in the sheared soft glassy materials to probe their structure in the direction orthogonal to shear. More precisely, our experiments (see Methods) consist in shearing the material horizontally in a Couette geometry and observing the possible vertical motion of solid particles under the action of gravity, thanks to a density mismatch $\Delta\rho$ between the particles and the material (Fig.1a). For small $\Delta\rho$, when the fluid is at rest the beads do not move, the gravity minus buoyancy force is not sufficient to break the jammed structure[19]. In contrast, when we shear the fluid and unjam it in the horizontal plane, the beads start moving perpendicularly at a constant velocity $V$ (Fig.2, Supplementary Fig.1).

The first important result is that the sedimentation of particles that were stable at rest is induced by shear: the orthoradial shear unjams the system in both the orthoradial and the vertical direction whatever the shear intensity. Moreover, we observe that the sedimentation velocity strongly increases, i.e. the drag force decreases, when the shear rate in the main direction increases (Fig.2); it also increases with the particle diameter and when the yield stress is decreased (Fig.2). In other words, the material seems to behave like a liquid with no yield resistance to flow in the vertical direction, and that is less and less viscous as it moves away from the jamming transition.

To understand these features, we performed experiments in more controlled conditions. We developed a new test (see Methods) in which we impose at the same time a shear flow and a squeeze flow (Fig.1b), and measure the material's resistance to both flows. These two flows involve independent components of the strain rate tensor $d_{ij}$, so that we can characterize in detail the response to a secondary flow. The material is set up between two parallel disks. The rotation of the upper disk around its axis induces a simple orthoradial shear flow of characteristic shear rate $\dot{\gamma} = 2d_{z\theta}$ and shear stress $\tau \equiv \tau_{z\theta}$. A squeeze flow is superimposed by moving the upper disk towards the other at controlled velocity; it induces mainly a simple radial shear flow of characteristic shear rate $\dot{\Gamma} = 2d_{rz}$ and shear stress $\Sigma \equiv \tau_{rz}$.

Our experiments start by imposing a decreasing rotational shear rate ramp without squeezing motion ($\dot{\Gamma} = 0$). In that case we obtain the typical flow curve of soft glassy materials in simple shear, namely a shear thinning behaviour tending to a plateau at low shear rates (Fig.3). The plateau level corresponds to the yield stress below which the material remains jammed. Then we start again the test, but now we impose an additional squeeze flow at constant squeeze shear rate $\dot{\Gamma} \neq 0$. In that case the flow curve is superimposed to the previous one (without squeezing) at high shear rates only i.e. when the rotational shear flow is dominant ($\dot{\Gamma} \ll \dot{\gamma}$, Regime 1) (Fig.3). At low shear rates, when the squeeze flow becomes dominant ($\dot{\Gamma} \gg \dot{\gamma}$, Regime 2), the new flow curve starts to depart from the previous one and the shear stress decreases strongly below the yield stress. A remarkable result is that in this regime the flow curve tends to follow a line of slope 1 in logarithmic scale at any observable rate, typical of a Newtonian behaviour (Fig.3). Moreover the viscous regime occurs for lower $\dot{\gamma}$ and is characterized by higher apparent viscosity when $\dot{\Gamma}$ is decreased. Similar features (not shown) were observed when studying the resistance to the squeeze flow: when the shear flow is dominant, the resistance to the squeeze flow is that of viscous fluid. This observation is critical as it further supports the generality of the subsequent theoretical analysis, independently of some specificity of the flows. In addition, we checked that these effects were not due to an experimental artifact such as wall slip.

This experiment, in which both superimposed flows are well controlled and characterized, now allows us to address the question of the quantitative resistance to flow. Our starting point is that the viscous resistance to the secondary flow decreases when the shear rate $\dot{\Gamma}$ of the main flow increases. This leads us to propose a structural interpretation, in analogy with the behaviour of liquids and glasses. The main flow introduces a structural relaxation timescale $1/\dot{\Gamma}$. Any secondary flow at a shear rate $\dot{\gamma} < \dot{\Gamma}$ then probes longer timescales, allowing for full relaxation of the material during this secondary flow which thus has a viscous behaviour. For $\dot{\gamma} > \dot{\Gamma}$ relaxation cannot take place and the nonlinear glassy behaviour is recovered. By analogy with simple liquids[20], we suggest that the apparent viscosity $\eta$ in the viscous regime is proportional to the relaxation timescale $1/\dot{\Gamma}$ and to a reference stress; a natural reference stress is that associated with the liquid/solid transition ($\tau_c$). This yields the viscosity of the secondary flow in the form $\eta \propto \tau_c/\dot{\Gamma}$. In the inset of Fig.3, we plot the experimentally measured effective viscosity of the rotational flow $\eta = \tau/\dot{\gamma}$ in Regime 2, scaled by $\tau_c$, vs. $\dot{\Gamma}$. Remarkably, these data are well fitted to the $1/\dot{\Gamma}$ curve. Note that due to the complex flow involved (see Methods) the factor close to 1 here observed is only due to our specific choice of the (constant) factors in the characteristic shear rate expressions. Anyway, this shows that, close to the jamming transition, the viscous resistance to the secondary flow is simply controlled by the apparent viscosity experienced in the main flow, i.e. when $\dot{\Gamma} \gg \dot{\gamma}$ we have $\tau \propto \eta(\dot{\Gamma})\dot{\gamma}$.

These results now allow analyzing quantitatively the sedimentation experiments: they suggest that the vertical drag force is that for a sphere moving through a Newtonian fluid of viscosity $\eta$ equal to that experienced by the fluid in the (dominant) horizontal simple shear. To test this prediction, we scale the measured sedimentation velocity by $V_{Newt.} = 2\Delta\rho g R^2/9\eta$, the velocity of a single sphere steadily falling through a Newtonian medium under the action of gravity, and we take $\eta = \tau(\dot{\gamma})/\dot{\gamma}$. Close to the jamming transition, all the $V/V_{Newt.}$ data fall around the same value (Fig.2 inset), in excellent agreement with the theory. The obtained ratio, 1.4, is higher than expected for a Newtonian fluid (around 0.7 for a 5% suspension), but this likely reflects some additional complexity of the flow field around several spheres moving through a complex fluid[21]. The collapse of the data obtained for various values of $\Delta\rho$, $R$ and $\eta$, proves the general validity of the suggested scaling, which confirms that the spheres basically "see" a simple viscous material in the direction orthogonal to shear.

These results make it possible to deduce the form of the 3D constitutive equation of soft glassy systems. In the liquid regime, the material has a similar apparent viscosity in the different directions when the flow in a given direction is dominant. This suggests that the stress tensor under such conditions takes a form analogous to that for a Newtonian fluid: $\tau_{ij} = 2\eta d_{ij}$, in which $d_{ij} = 1/2\left(du_i/dx_j + du_j/dx_i\right)$ is the strain rate tensor ($u_i$ is the material velocity in the direction $x_i$). In contrast, with Newtonian fluids $\eta$ is not constant but depends on the flow intensity in the main direction. Under simple shear, it reads $\eta = \left(\tau_c + k\dot{\gamma}^n\right)/\dot{\gamma}$. We can extrapolate this equation to find the 3D constitutive equation of soft glassy systems in their liquid regime, by using in the apparent viscosity, instead of the shear rate, a generalized shear intensity $d = \sqrt{2\sum_{i,j} d_{ij}^2}$ which is an invariant of the strain rate tensor to ensure frame invariance:

$$\tau_{ij} = 2\frac{\left(\tau_c + kd^n\right)}{d} d_{ij} \qquad (1)$$

This expression is in good agreement with our observations: it predicts that for a dominant squeeze flow, near the jamming transition, the resistance to the rotational shear is characterized by a purely viscous stress that reads $\tau = \tau_{z\theta} = \frac{\tau_c}{\Gamma}\dot{\gamma}$ (with $\dot{\gamma} = 2d_{z\theta}$).

The constitutive equation of soft glassy systems also includes a yielding criterion associated with unjamming, which marks the transition from the solid to the liquid regime. In a pure simple shear this transition occurs when the shear stress $\tau$ becomes larger than a single scalar $\tau_c$ (the yield stress), but the above data show that as soon as the material is unjammed in one direction it is unjammed in another direction. This suggests that the 3D criterion for unjamming involves some distance of the stress tensor from a critical value. In this context, the most natural approach consists in using for this distance the sum of the squares of the shear stress components $\sqrt{0.5\sum_{i \neq j} \tau_{ij}^2} = \tau_c$, which can be written more consistently as a function of an invariant of the stress tensor to include possible normal force differences:

$$\sqrt{0.5 \sum_{i,j} \tau_{ij}^2} = \tau_c \qquad (2)$$

which is the Von Mises criterion commonly used for solid materials[14]. In simple shear and simple squeeze, this criterion gives back the usual yielding criterion ($\tau = \tau_{z\theta} < \tau_c$, $\Sigma = \tau_{rz} < \tau_c$). We have tested further the validity of this 3D yield criterion by combining squeeze and rotational shear flows and measuring simultaneously the ($\tau_{rz}, \tau_{z\theta}$) values at the onset of flow. The corresponding data are presented in Fig. 4. The jammed region effectively lies within the disk defined by $\sqrt{\tau_{rz}^2 + \tau_{z\theta}^2} < \tau_c$, in good agreement with Eq. 2. It is remarkable that the same plasticity criterion is found in soft glassy materials as in most usual crystalline solids[14], while the microscopic plasticity mechanisms seem radically different[5]. However, a yield criterion implying $\tau_{rz}^2 + \tau_{z\theta}^2$ is not unexpected from a pure *rational mechanics*[22] point of view under the (unobvious) hypothesis that the material is isotropic at yield (which implies that the yielding criterion is a function of the three invariants of the stress tensor[23]) and in the absence of normal stress differences.

Finally, Eqs. (1) and (2) constitute a 3D modelling of the jamming and flows of soft glassy systems deduced from experiments. Beyond their obvious interest for complex flow modelling these results provide generic quantitative explanations for shear-induced heterogeneities in industrial flows of suspensions in yield stress fluid, and for geophysical problems as the liquefaction of quicksand under shear[24]. The generality of this constitutive equation is supported by the fact that we observed the same trends in sedimentation and squeeze-shear tests for our soft glassy materials (emulsions, bentonite suspensions, Carbopol gels) of different structures. This suggests that the main trends here observed may find a same phenomenological origin in all materials, namely the shear-induced structural relaxation of their structural units.

Our macroscopic observations actually shed light on the microscopic behaviour of such materials. It has been observed on several soft glassy systems[4,25,26,27] that, under shear, in addition to cage diffusion observed at rest[2,3,28], the elements undergo a simple diffusion process as in simple liquids in the three directions of space; other, consistent, observations are that the sheared colloidal glasses experience full exponential alpha relaxation in the directions orthogonal to shear[4] as in a liquid state. Our results provide the missing macroscopic counterpart to these microscopic observations: if the Stokes-Einstein relation applies[25], the particles should indeed diffuse as in a simple liquid of viscosity $\eta$ equal to that we observe in the direction orthogonal to shear, and the diffusion coefficient $D$ should be proportional to $1/\eta$. Then, in the low shear rate limit, our approach predicts that $D \propto \dot{\gamma}$. This is in quantitative agreement with some observations[25,26]; the nonlinear scaling of D with $\dot{\gamma}$ observed by others[4,27], consistent with the alpha relaxation time measurements[4] and with theoretical predictions[29], may be due to the fact that the diffusion coefficient considered in these studies is that of the materials constitutive elements, which do not probe the macroscopic behaviour.

Our results finally show with new quantitative results that there are striking similarities between the structural relaxation driven by shear, temperature, and density at the approach of the jamming transition. The flow curves of Fig.3 are actually remarkably similar to those observed in colloidal suspensions[15,30] and predicted in mode coupling theories[16,30] (MCT) when varying the volume fraction $\phi$ of colloids near that of the colloidal glass transition $\phi_g$. When $\phi < \phi_g$, colloidal suspensions have a Newtonian behaviour below a low critical shear rate $\dot{\gamma}_c(\phi)$, whereas they have the same shear-thinning behaviour as the colloidal glass at higher shear rates[15,16,30]. This Newtonian behaviour is observed when the timescale for the diffusion of the particles out of their cages is smaller than the typical flow timescale[16] $1/\dot{\gamma}$. As a consequence, $\dot{\gamma}_c$ decreases strongly as $\phi$ approaches $\phi_g$, similarly to our observations when varying the squeeze shear rate $\dot{\Gamma}$. Similarly, a high viscosity plateau is found at lower and lower shear rates when the temperature of metallic glasses is decreased[31]. Overall, MCT thus seems to be a very good candidate for general modelling of the jamming transition in all glassy systems, and to account properly for the effect of shear. In particular, it is remarkable that our finding of a Von Mises jamming criterion is consistent with very recent MCT findings[32]. However, one still has to test MCT predictions in view of our observations and of the 3D constitutive law we propose.

METHODS

MATERIALS PREPARATION
The emulsions used in the shear-induced sedimentation experiments are prepared by dispersing a 100 g/l water solution of $CaCl_2$ in a solution of *Span* 80 emulsifier (7%) in dodecane oil at 6000 rpm with a Silverson L4RT mixer. The emulsions used in the combined squeeze and rotational shear flows experiments are prepared by dispersing dodecane oil at 5000 rpm in a 2 g/L water solution of tetradecyltrimethylammonium Bromide (TTAB). The droplet size is 1 micron. Seven emulsions are used in this study; their droplet concentration is between 72 and 85% and their yield stresses range between 8.5 and 52 Pa.

The physical gel is prepared by dispersing Carbopol 980 (Noveon) in water at a 0.4% concentration and neutralized with NaOH at pH=7. Carbopol gels may then be seen as polydisperse glasses made of individual swollen hydrophilic elastic sponges. The Carbopol gel used in this study has a 70 Pa yield stress.

The colloidal gel is a suspension of 3.5% Na-Bentonite particles in water. Bentonite is a natural swelling clay with slightly flexible, large aspect ratio particles (typical length: 1 μm, typical thickness: 0.01 μm) which can aggregate via edge-to-face links, so that the suspension is a colloidal gel with a thixotropic yield stress fluid behaviour. The suspension was prepared by a strong mixing of the solid phase with water, then left at rest three months before any test, which avoids further irreversible (chemical) aging over the duration of the experiments. The bentonite suspension used in this study has a 4 Pa dynamic yield stress. While bentonite suspensions are thixotropic yield stress materials, time effects during flows are negligible for the suspensions we study and for the time scales of our experiments. The flowing suspension can be then considered as a simple yield stress fluid with a (dynamic) yield stress associated with stoppage for a progressive stress decrease. On the other hand, aging effects are important at rest and would imply additional complexity for determining a yield criterion with the (static) yield stress associated with flow start.

SHEAR-INDUCED SEDIMENTATION EXPERIMENT

Monodisperse glass beads of density 2.5, of diameter between 140 and 420 microns (with 10% polydispersity) are suspended in yield stress fluids at a 5% volume fraction to avoid important collective effects. For all materials at rest, the external force exerted on the particles, i.e. the gravity force minus the buoyancy force $F_g = 4\pi R^3 g \Delta\rho/3$, is much smaller than the critical resistance to motion[33] $F_c \approx 14\pi R^2 \tau_c$ (it is of order of 1% of $F_c$ in all cases).

The material is loaded in a Couette geometry, the dimensions of which are: inner cylinder radius 4.1 cm, outer cylinder radius 6cm and inner cylinder height 11 cm. Sandpaper of roughness equivalent to the particles size is glued on the walls to avoid wall slip. The suspension is sheared at constant macroscopic shear rate, and the particles vertical velocity is obtained thanks to the evolution in time of the spatial distribution of the particle concentration measured through Magnetic Resonance Imaging (MRI) techniques. The MRI set up was described in detail elsewhere[34]. The particle volume fraction can be obtained both in the radial and vertical directions from density imaging[34] with an accuracy of 0.3%. Since the shear distribution within the gap of a Couette geometry is heterogeneous it is crucial to impose a large rotation velocity to avoid shear localization. Moreover, we measure the vertical concentration profiles in a 0.8mm thick layer in the middle of the gap to obtain information on particle motion in a zone of approximately homogeneous shear rate. This last shear rate is obtained locally through MRI thanks to the velocity profile measurement[34].

At rest, the particles appear to remain indefinitely in their initial position (Supplementary Fig.1 inset): there is no observable difference between the vertical concentration profiles measured after loading and after a 24 h rest.

Then we start shearing the materials by rotating the inner cylinder at a given rotation velocity $\Omega$ between 20 and 130 rpm (corresponding to a shear rate ranging from 4 to 25 $s^{-1}$). While the particles are stable at rest, we observe that there is sedimentation when the material is sheared. The sedimentation profiles show classical features of sedimentation in Newtonian fluids (Supplementary Fig.1): the upper part is at a 0% concentration; the bottom part remains at the initial 5% concentration (note that the particles tend to accumulate out of the measurement window, below the inner cylinder, which explains that the concentration in the lower part of our profiles does not change); the transition zone between these two parts is rather narrow (with a typical thickness of 6mm), and defines a sedimentation front that moves regularly towards the bottom as the flow duration increases. We observe that this front moves linearly in time (Supplementary Fig.1), defining a constant front velocity $V$ that is identified with the individual particles sedimentation velocity.

COMBINED ROTATIONAL SHEAR AND SQUEEZE FLOWS EXPERIMENTS

A constant volume of material (typically 0.7 ml) is inserted within the gap of a parallel disk geometry of a Kinexus rheometer (Malvern Instruments) so as to partially fill the gap (initial height $h_0$=1.1 mm). The material is then sheared by rotating the upper disk around its axis at a rotational velocity $\Omega$, and at the same time it is squeezed by moving the upper disk towards the other at controlled velocity $V$ (Fig. 1). We record the torque $T$ exerted onto the axis of the upper plate and the normal force $F$ exerted on the upper plate. Due to the squeeze flow, the material geometric characteristics change in time. In the following analysis, the relationships between the shear stresses and shear rates and the macroscopic quantities thus involve the instantaneous material outer radius $R(t)$ and thickness $h(t)$.

During this test, the flow characteristics within the gap are somewhat heterogeneous but as usual in viscometry it is possible to follow average variables which correctly reflect the relative variations of the local variables[23]. The applied torque induces an orthoradial motion associated with the velocity $v_\theta$ (in the cylindrical frame shown in Fig.1b) linked to the stress component $\tau_{z\theta}$. The resulting shear rate due to the vertical gradient of $v_\theta$ scales with $\dot{\gamma} = \Omega R/h$ and from the momentum balance it is found that the shear stress scales with $\tau = T/R^3$. Moreover, it may be shown that there is a relationship between $\tau$ and $\dot{\gamma}$ via a function depending only on the intrinsic material behavior[23]: these variables thus play the role of reference shear stress and shear rate the relative variations of which provide the characteristics of the material behavior.

In a similar way, the applied normal force is at the origin of a radial pressure gradient which induces mainly a radial shear flow[13] (as long as $h/R \ll 1$) associated with the velocity $v_r$, linked to a stress component $\tau_{rz}$. From mass conservation (which provides the relationship between the vertical velocity $V$ and the average radial velocity) it is found that the resulting shear rate due to the vertical gradient of $v_r$ scales with $\dot{\Gamma} = VR/h^2$, and from the momentum balance it is found that the shear stress scales with $\Sigma = Fh/R^3$. Again, it may be shown that there is a relationship between $\Sigma$ and $\dot{\Gamma}$ via a function depending only on the intrinsic material behavior[23]. Thus, in this paper, we simply follow these quantities, namely $\tau$ and $\dot{\gamma}$ for the rotational shear flow and $\Sigma$ and $\dot{\Gamma}$ for the squeeze flow. Note finally that the critical stresses for solid-liquid transition are here associated with the stress plateaus observed beyond a critical deformation for very low flow rates[11]. Under these conditions, there are no radial stress heterogeneity issues since at the stress plateau the whole material is at yield. For the consistency of the mechanical characterization of our materials, we thus chose to use $\tau = 3T/2\pi R^3$ and $\Sigma = 3Fh/2\pi R^3$, which can be shown to provide the exact yield stress of a plastic material respectively in simple shear and simple squeeze[11,35] (for $h/R \ll 1$).

---

1  Liu, A. J. & Nagel, S. R. (eds) *Jamming and Rheology: Constrained Dynamics on Microscopic and Macroscopic Scales.* Taylor & Francis, New York (2001).

2. Kegel, W. K & van Blaaderen, A. Direct observation of dynamical heterogeneities in colloidal hard-sphere suspensions. *Science* **287**, 290-293 (2000).

3. Weeks, E. R., Crocker, J. C., Levitt, A. C., Schofield, A. & Weitz, D. A. Three-dimensional direct imaging of structural relaxation near the colloidal glass transition. *Science* **287**, 627-631 (2000)

4. Besseling, R., Weeks, E. R., Schofield, A. B., & Poon W. C. K. Three-dimensional imaging of colloidal glasses under steady shear. *Phy. Rev. Lett.* **99**, 028301 (2007).

5. Schall, P., Weitz, D. A. & Spaepen, F. Structural rearrangements that govern flow in colloidal glasses. *Science* **318**, 1895-1899 (2007).

6. Ballesta, P., Duri, A. & Cipelletti, L. Unexpected drop of dynamical heterogeneities in colloidal suspensions approaching the jamming transition. *Nature Phys.* **4**, 550-554 (2008).

7. Liu, A. J. & Nagel, S. R. Jamming is not just cool any more. *Nature* **396**, 21–22 (1998).

8. Pusey, P. N. & van Megen, W. Observation of a glass-transition in suspensions of spherical colloidal particles. *Phys. Rev. Lett.* **59,** 2083-2086 (1987).

9. Sollich, P., Lequeux, F., Hébraud, P. & Cates, M. E. Rheology of Soft Glassy Materials. *Phys. Rev. Lett.* **78**, 2020-2023 (1997).

10. Bonn, D. & Denn, M. Yield stress fluids slowly yield to analysis. *Science* **324**, 1401-1402 (2009).

11. Coussot, P. *Rheometry of Pastes, Suspensions and Granular Materials*. John Wiley & Sons, New York (2005).

12. Nguyen, Q. D. & Boger, D. Measuring the flow properties of yield stress fluids. *Annual Review of Fluid Mechanics* **24**, 47-88 (1992).

13. Coussot, P. Rheophysics of pastes: a review of microscopic modelling approaches. *Soft Matter* **3**, 528–540 (2007).

14. Hill, R. *The Mathematical Theory of Plasticity*. Oxford University Press, Oxford (1998).

15. Petekidis, G., Vlassopoulos, D. & Pusey, P. Yielding and flow of sheared colloidal glasses. *J. Phys. Cond. Matt.* **16**, 3955-3964 (2004).

16. Fuchs, M. & Cates, M. E. Theory of nonlinear rheology and yielding of dense, colloidal suspensions. *Phys. Rev. Lett.* **89**, 248304 (2002).

17. Goyon, J., Colin, A., Ovarlez, G., Ajdari, A. & Bocquet, L. Spatial cooperativity in soft glassy flows. *Nature* **454**, 84-87 (2008).

18. Lemaître, A & Caroli, C. Rate-Dependent Avalanche Size in Athermally Sheared Amorphous Solids. *Phys. Rev. Lett.* **103**, 065501 (2009).

19. Tabuteau, H., Coussot, P. & de Bruyn, J.R. Drag force on a sphere in steady motion through a yield-stress fluid. *J. Rheol.* **51**, 125–137 (2007).

20. Jones, R. A. L. *Soft Condensed Matter*. Oxford University Press, Oxford (2002).

21. Chhabra R. P. *Bubbles, Drops And Particles in Non-Newtonian Fluids, Second Edition*. CRC Press, Boca Raton (2007).

22  Truesdell, C. *A First Course in Rational Continuum Mechanics*. Pure and Applied Mathematics, Academic Press, New York, San Francisco, London (1977).

23  Coleman, B.D., Markovitz, H. & Noll, W. *Viscometric Flows of Non-Newtonian Fluids; Theory and Experiment.* Springer, Berlin, New York (1966).

24  Khaldoun, A., Eiser, E., Wegdam, G. H. & Bonn, D. Rheology: Liquefaction of quicksand under stress. *Nature* **437**, 635–635 (2005).

25  Ono, I., O'Hern, C., Durian, D., Langer, S., Liu, A. & Nagel, S. Effective temperatures of a driven system near jamming. *Phys. Rev. Lett.* **89**, 95703 (2002).

26  Song, C., Wang, P. & Makse, H. A. Experimental measurement of an effective temperature for jammed granular materials. *PNAS* **102**, 2299-2304 (2005).

27  Möbius, M. E., Katgert, G. & van Hecke, M. Relaxation and flow in linearly sheared two-dimensional foams. preprint arXiv:0811.0534v2 (2009).

28  Oppong, F. K., Coussot, P. & de Bruyn, J. R. Gelation on the microscopic scale. *Phys. Rev. E* **78**, 021405 (2008).

29  Saltzman, E. J., Yatsenko, G. & Schweizer, K. S. Anomalous diffusion, structural relaxation and shear-thinning in glassy hard sphere fluids. *J. Phys.: Condens. Matter* **20**, 244129 (2008).

30  Fuchs, M. & Ballauff, M. Flow curves of dense colloidal dispersions: Schematic model analysis of the shear-dependent viscosity near the colloidal glass transition. *J. Chem. Phys.* **122**, 094707 (2005)

31  Lu, J., Ravichandran, G. & Johnson, W. L. Deformation behavior of the Zr41.2Ti13.8CU12.5Ni10Be22.5 bulk metallic glass over a wide range of strain-rates and temperatures. *Acta Materialia* **51,** 3429-3443 (2003).

32  Brader, J. M., Voigtmann, T., Fuchs, M., Larson, R. G. & Cates, M. E. Glass rheology: From mode-coupling theory to a dynamical yield criterion. *PNAS,* doi:10.1073/pnas.0905330106 (2009).

33  Beris, A.N., Tsamopoulos, J.A., Armstrong, R.C. & Brown, R.A. Creeping motion of a sphere through a Bingham plastic. *J. Fluid Mech.* **158**, 219-244 (1985).

34  Ovarlez, G., Bertrand, F. & Rodts, S. Local determination of the constitutive law of a dense suspension of noncolloidal particles through magnetic resonance imaging. *J. Rheol.* **50**, 259 (2006).

35  Rabideau, B. D., Lanos, C. & Coussot, P. An investigation of squeeze flow as a viable technique for determining the yield stress. *Rheol. Acta* **48**, 517-526 (2009).

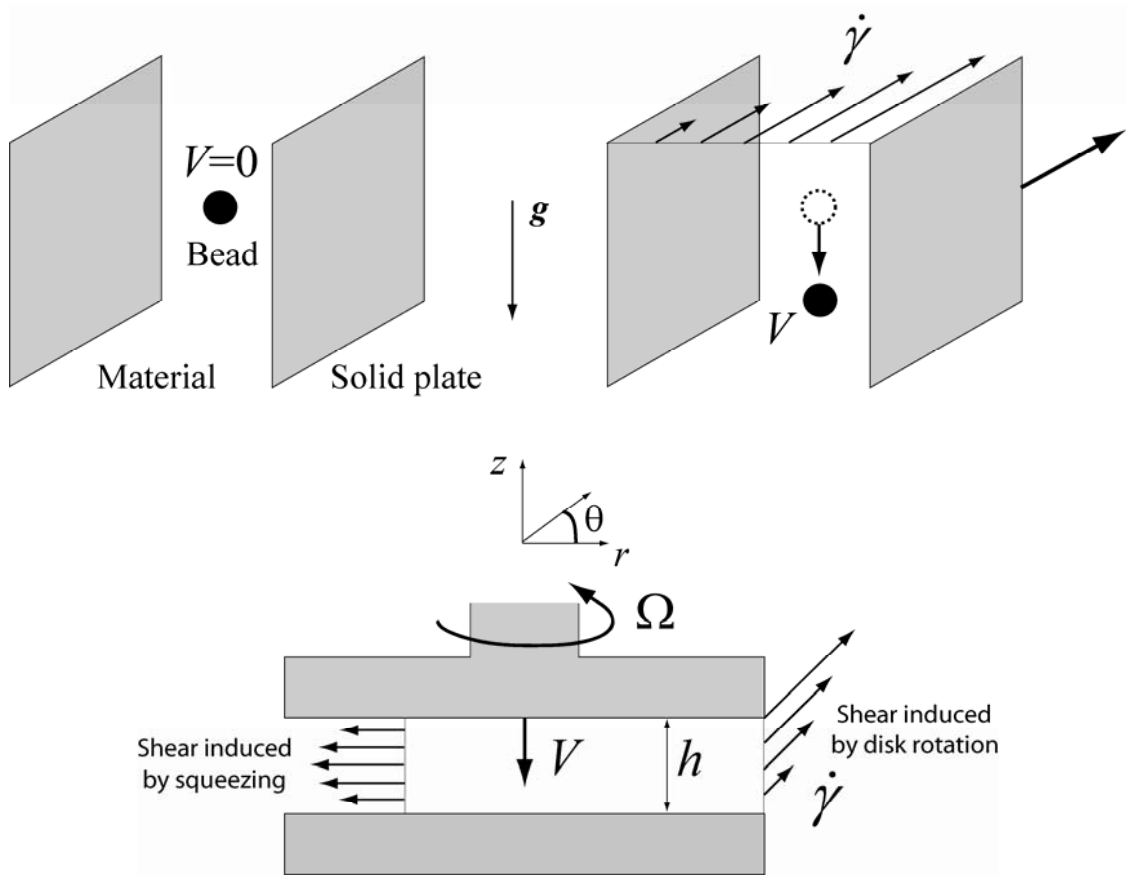

**Figure 1 | Sketches of the experiments.** a) Sketch of the sedimentation experiment perpendicular to a shear flow: (left) bead suspended in the material at rest; (right) bead falling under the action of gravity in the material sheared in the horizontal plane. b) Sketch of the combined squeeze and rotational shear flow experiments: the material is loaded between two parallel plates, and is simultaneously sheared by rotating the upper disk around its axis at a rotational velocity $\Omega$, and squeezed by moving the upper disk towards the other at controlled velocity $V$.

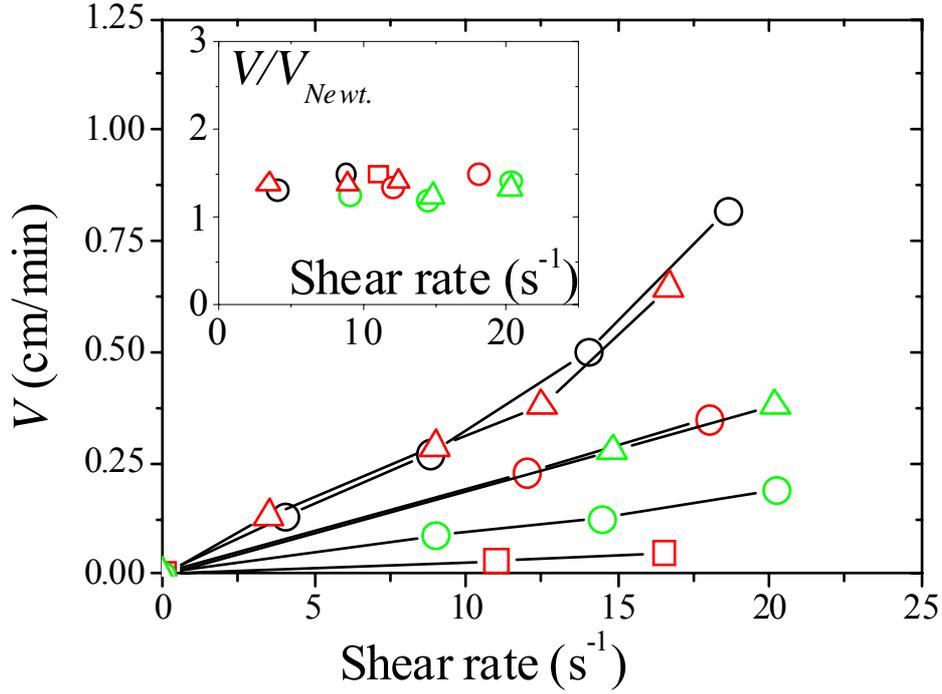

**Figure 2 | Shear-induced sedimentation velocity.** Sedimentation velocity $V$ of glass beads in a sheared emulsion as a function of the applied shear rate $\dot{\gamma}$, for various bead diameter $d$ and various emulsion yield stresses $\tau_c$ (squares: $d$=140 µm, $\tau_c$=25 Pa; circles: $d$=275 µm, $\tau_c$=8.5 Pa (black), 15 Pa (red), 33 Pa (green); triangles: $d$=400 µm, $\tau_c$=21.5 Pa (red), 33 Pa (green)). Inset: same data close to the jamming transition, rescaled by the velocity $V_{Newt.} = 2\Delta\rho g R^2/9\eta$ of a single sphere that would fall in a viscous medium of viscosity $\eta = \tau(\dot{\gamma})/\dot{\gamma}$. Note that in some systems sedimentation velocities at the highest shear rates tested are larger than the plateau value evidenced here, an effect likely due to non-Newtonian effects (such as normal stresses) on sedimentation[21], which is out of the scope of our study.

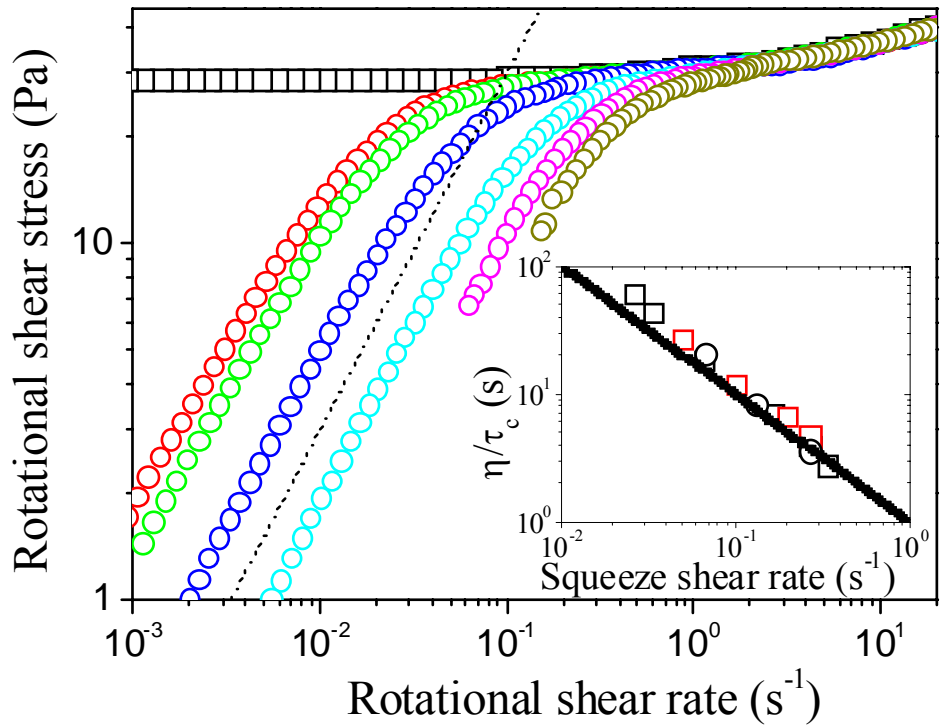

**Figure 3 | 3D flow curve.** Rotational shear stress vs. rotational shear rate measured in a 28 Pa yield stress emulsion by rotating the upper plate of a parallel plate geometry, for a constant gap (squares) and while simultaneously squeezing the material (circles) at various squeeze shear rates $\dot{\Gamma}$ (from left to right: $\dot{\Gamma} = 0.027\,\text{s}^{-1}$ (red), $\dot{\Gamma} = 0.033\,\text{s}^{-1}$ (green), $\dot{\Gamma} = 0.067\,\text{s}^{-1}$ (blue), $\dot{\Gamma} = 0.167\,\text{s}^{-1}$ (cyan), $\dot{\Gamma} = 0.267\,\text{s}^{-1}$ (pink), $\dot{\Gamma} = 0.333\,\text{s}^{-1}$ (brown)). The dotted line is a viscous law. Inset: Viscosity $\eta$ of the rotational flow in the low rotational shear rate regime scaled by the yield stress $\tau_c$ of the materials (squares: 28 Pa (black) and 15 Pa (red) yield stress emulsions, circles: 3.9 Pa yield stress bentonite suspension) vs. the squeeze shear rate $\dot{\Gamma}$; the line is a $1/\dot{\Gamma}$ function.

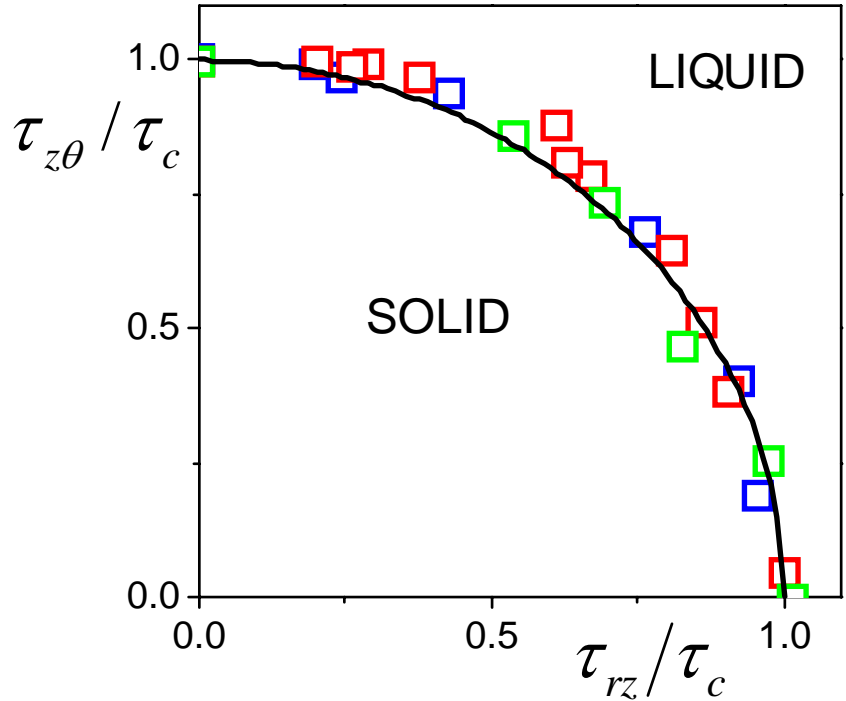

**Figure 4 | 3D yield criterion.** Jamming phase diagram in a rotational shear stress $\tau_{z\theta}$ vs. squeeze shear stress $\tau_{rz}$ plane, scaled by the yield stress $\tau_c$ measured in simple shear. The squares are the experimentally measured stresses at the onset of flow when combining rotational shear and squeeze shear flows, for two different emulsions (blue: $\tau_c$=28 Pa, red: $\tau_c$=52 Pa) and a Carbopol gel (green, $\tau_c$=70 Pa). The line is the Von Mises criterion $\sqrt{\tau_{z\theta}^2 + \tau_{rz}^2} = \tau_c$.

# Supplementary Information

*Three-dimensional jamming and flows of soft glassy materials*

G. Ovarlez, Q. Barral, P. Coussot

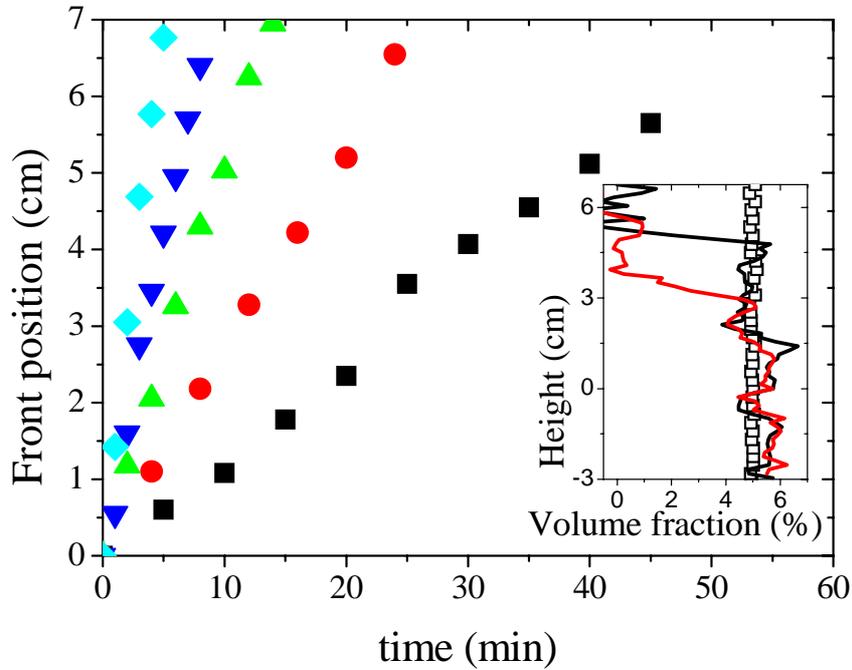

**Supplementary Figure 1 | Bead sedimentation observed through MRI techniques.** Position of the sedimentation front as a function of the time of shear, for a 5% suspension of 275 microns glass beads in an emulsion of 8.5 Pa yield stress, for various shear rates: 4 s$^{-1}$ (squares), 8.8 s$^{-1}$ (circles) 14 s$^{-1}$ (up triangles) 18.6 s$^{-1}$ (down triangles) 25 s$^{-1}$ (diamonds). Inset: vertical volume fraction profiles observed in the gap of the Couette geometry in the same material as in Fig.2, after a 24 h rest (squares) and after 15 min (black line) and 25 min (red line) of shear at 4 s$^{-1}$.